# TECHNICAL DESIGN STUDIES OF TAC SASE FEL PROPOSAL*

B. Ketenoglu#, O. Yavas, M. Tural, Ankara University, Ankara, Turkey
S. Ozkorucuklu, S. Demirel University, Isparta, Turkey
I. Tapan, O. Sahin, Uludağ University, Bursa, Turkey
P. Arikan, Gazi University, Ankara, Turkey

*Abstract*

A SASE FEL facility was first proposed in Feasibility Report of the TAC (Turkish Accelerator Center) project in 2000. Conceptual Design Report (CDR) of the project was completed in 2005. Technical Design Report (TDR) studies of TAC were started in 2006 in frame of an inter universities project with support of State Planning Organization (SPO) of Turkey. Main goal of the SASE FEL proposal is to cover VUV and soft X-rays region of the spectrum besides IR-FEL, Bremsstrahlung and Synchrotron Radiation proposals of TAC. Up to now, optimization studies based on a special RF linac or an Energy Recovery Linac (ERL) for the SASE FEL facility, were completed. Today, ERLs provide a powerful broad range of applications like: electron cooling devices, high average brightness, high power FELs, short-pulse radiation sources and high luminosity colliders. In this study, main parameters for two linac options and SASE FEL are given.

## OVERVIEW OF THE TAC SASE FEL FACILITY PROPOSAL

It was first planned that, TAC linac-ring type collider's 1 GeV electron linac should asynchronously be used as a driver for SASE FEL facility [1], as shown in Fig. 1. Therefore, SASE FEL optimization studies has been based on a 1 GeV superconducting RF linac [2] since the ERL idea appears for both the collider and SASE FEL proposals of the project. On the other hand, i.e. Fig. 2, takes a glance at the SASE FEL facilities around the World classified by their linac technology in a bird's eye view.

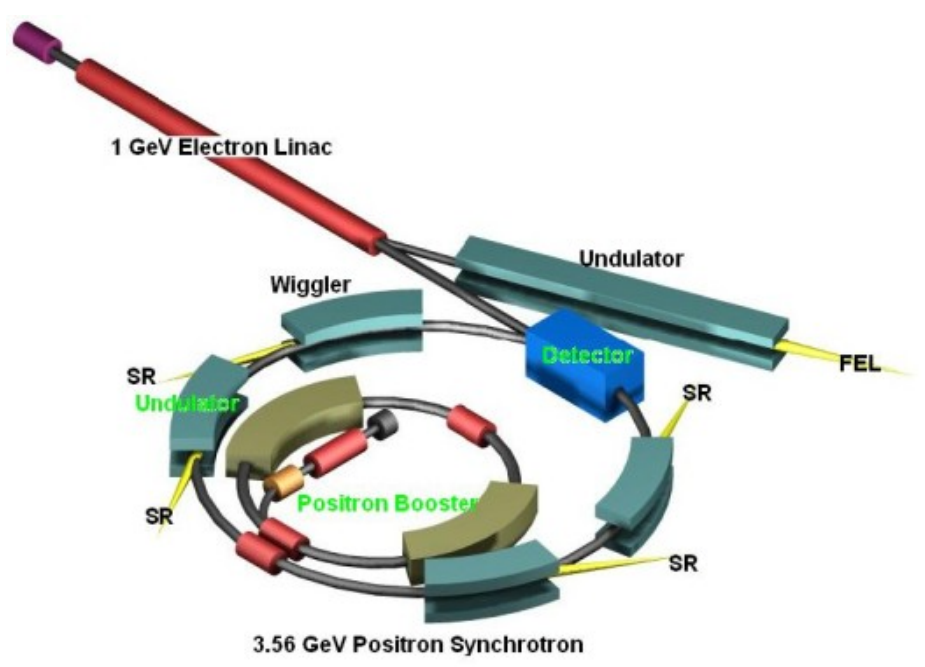

Figure 1: Schematic view of TAC linac-ring type $e^-e^+$ collider and SASE FEL facility.

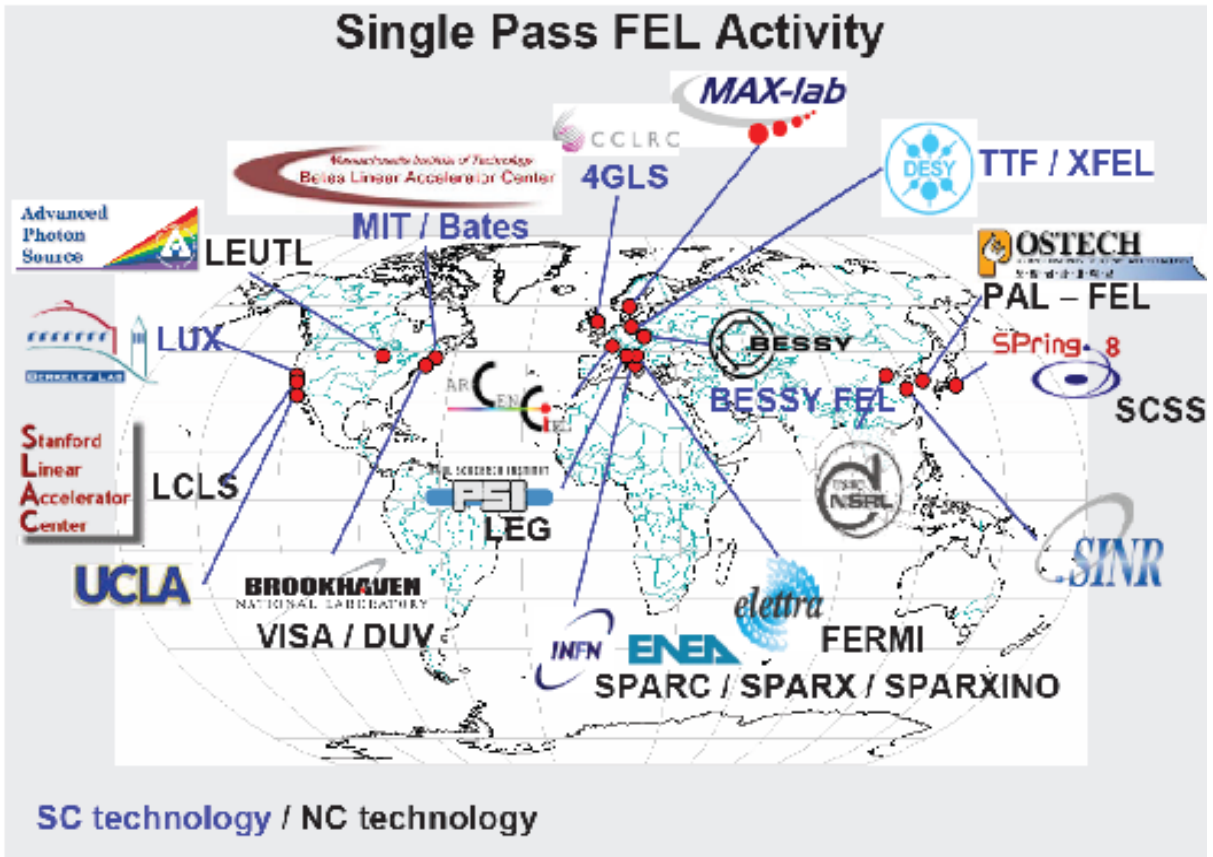

Figure 2: SASE FEL facilities classified by their linac technology around the World. The blues are superconducting and the blacks are normal conducting.

## THE ELECTRON SOURCE & LINAC

TAC 1 GeV superconducting linac was planned to be based on a photo-cathode RF gun. Electrons released from the gun, are accelerated in superconducting RF cavities and compressed by magnetic chicanes between the modules. After several diagnostic units, accelerated electron bunches (up to 1 GeV) pass through the undulator to achieve FEL process, as shown in Fig. 3.

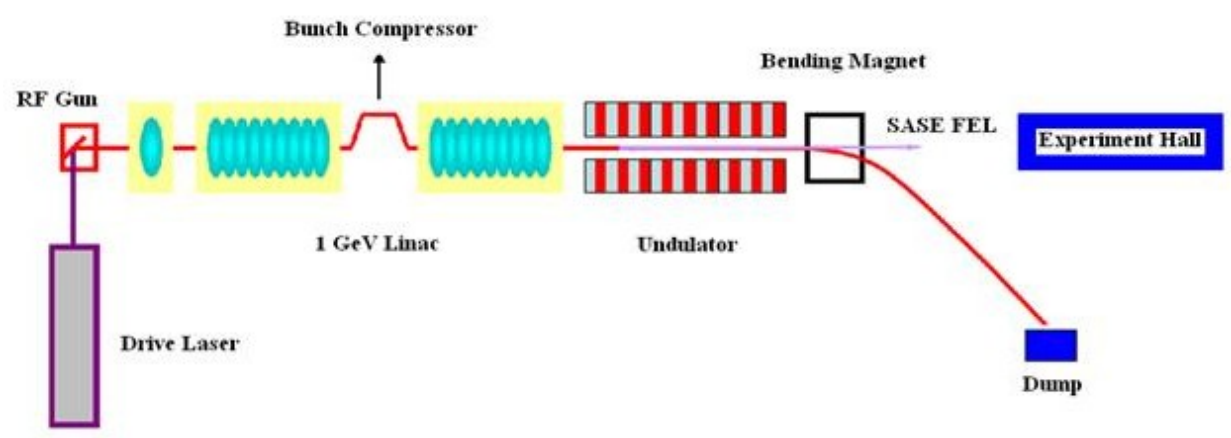

Figure 3: Schematic view of TAC SASE FEL facility

Proposed RF gun parameters and optimized tentative electron beam parameters [2] are given in Table 1. Definitely, the time structure of the electron beam will be in pulsed mode and in due course, L-Band (1.3 GHz) TESLA type superconducting RF cavities will probably be used.

*Work is supported by Turkish State Planning Organization
#bketen@eng.ankara.edu.tr

Table 1: Proposed Electron Beam Parameters Based On A Photo-Cathode RF Gun.

| Electron Beam | Value |
|---|---|
| Beam energy (GeV) | 1 |
| Number of electrons per bunch ($x10^9$) | 5.5 |
| Beam current (mA) | 26.4 |
| Peak current (A) | 2106 |
| Energy spread (%) | 0.1 |
| Normalized emittance (μm.rad) | 3.1 |
| Transverse beam sizes (μm) | 75.2 |
| Longitudinal bunch length (mm) | 0.05 |

## THE UNDULATOR & TENTATIVE SASE FEL PARAMETERS

Fundamentally, it is planned to cover 1-100 nm range of the spectrum. This condition directly restricts the transverse emittance of the electron beam between $0.08 < \varepsilon_t < 0.8$ nm range. On the other hand, optimizations on FEL process has been based on a Samarium Cobalt planar undulator [2] as shown in Fig. 4.

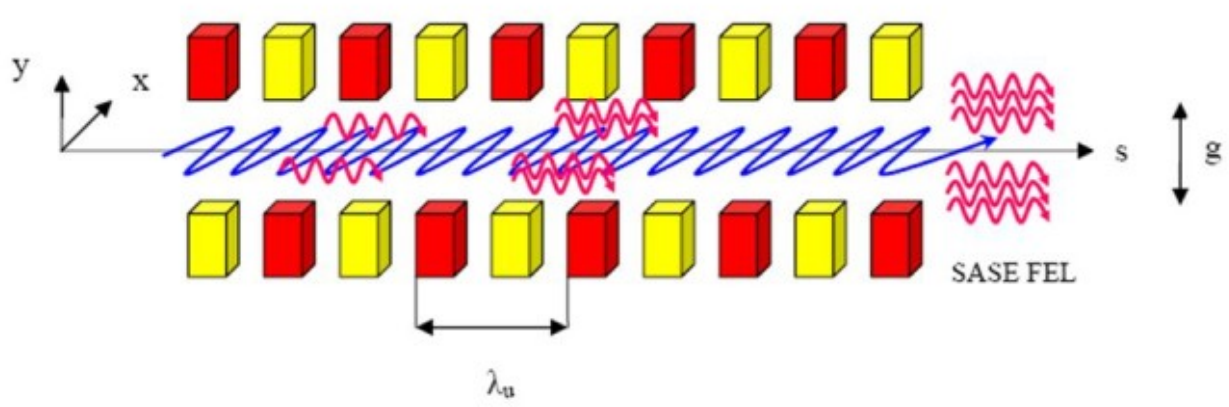

Figure 4: Schematic view of a planar undulator

The wavelength can easily be found by variying the undulator gap and period from, i.e. Fig. 5, certainly taking the technological limits into consideration. In Table 2, undulator gap is varied from 1 cm to 1.5 cm by 0.1 cm steps and undulator period is varied from 3 cm to 5 cm by 1 cm steps for each gap value. 1., 4., 5., 7., 8., 10., 11., 13., 14., 16. and 17. rows are convenient choices but the other exluding rows are inconvenient because of the K values, they are exceeding the K limit for undulators.

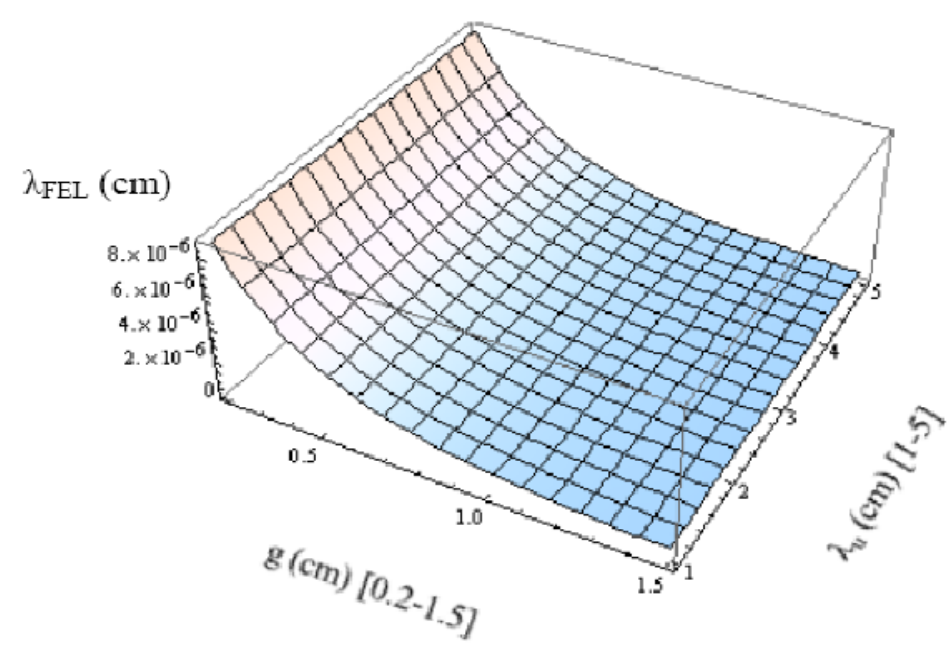

Figure 5: $\lambda_{FEL}$ vs g & $\lambda_u$

In Table 2, the seventh row from above seems more feasible, thus undulator optimization is based on dedicated row and the optimized undulator parameters [2] are given in Table 3.

Table 2: Convenient Parameter Choices For A Typical Samarium Cobalt Planar Undulator.

| g (cm) | $\lambda_u$ (cm) | g/ $\lambda_u$ | ~ B (T) | ~ K | ~ $\lambda_{FEL}$ (nm) |
|---|---|---|---|---|---|
| 1 | 3 | 0.33 | 0.657 | 1.84 | 10.5 |
| 1 | 4 | 0.25 | 0.949 | 3.55 | 38 |
| 1 | 5 | 0.2 | 1.198 | 5.6 | 108 |
| 1.1 | 3 | 0.37 | 0.571 | 1.6 | 9 |
| 1.1 | 4 | 0.275 | 0.848 | 3 | 32 |
| 1.1 | 5 | 0.22 | 1.1 | 5.14 | 90 |
| 1.2 | 3 | 0.4 | 0.498 | 1.4 | 7.7 |
| 1.2 | 4 | 0.3 | 0.759 | 2.84 | 26.2 |
| 1.2 | 5 | 0.24 | 0.994 | 4.64 | 76 |
| 1.3 | 3 | 0.43 | 0.436 | 1.22 | 6.8 |
| 1.3 | 4 | 0.325 | 0.68 | 2.54 | 22.1 |
| 1.3 | 5 | 0.26 | 0.91 | 4.25 | 66 |
| 1.4 | 3 | 0.47 | 0.384 | 1.07 | 6.2 |
| 1.4 | 4 | 0.35 | 0.612 | 2.28 | 18.8 |
| 1.4 | 5 | 0.28 | 0.829 | 3.87 | 55 |
| 1.5 | 3 | 0.5 | 0.339 | 0.95 | 5.7 |
| 1.5 | 4 | 0.375 | 0.551 | 2.05 | 16 |
| 1.5 | 5 | 0.3 | 0.758 | 3.54 | 47 |

Table 3: Undulator Parameters

| Undulator | Value |
|---|---|
| Period length (cm) | 3 |
| Gap (cm) | 1.2 |
| Peak magnetic field, $B_u$ (T) | 0.498 |
| K parameter | 1.395 |
| Saturation length (m) | 36 |
| Number of periods | 1200 |

Finally, laser parameters based on a superconducting RF linac driven Samarium Cobalt planar undulator [2], are given in Table 4.

Table 4: SASE FEL Parameters

| SASE FEL | Value |
|---|---|
| Wavelength (nm) | 7.7 |
| Photon energy (eV) | 160.5 |
| ρ parameter | 0.0018 |
| Peak power (GW) | 1.4 |
| Average power (kW) | 21.8 |
| Gain length, $L_g$ (m) | 0.75 |
| Gain length, 3D $L_g$ (m) | 1.57 |
| Peak flux (photons/s) | $1.5x10^{26}$ |
| Peak brightness (photons/s/mrad$^2$/0.1%bg) | $1.7x10^{29}$ |
| Peak brilliance (photons/s/mm$^2$/mrad$^2$/0.1%bg) | $2.9x10^{30}$ |

## A MORE PROMISING OPTION: ENERGY RECOVERY LINAC

A more promising option, namely ERL, is arising nowadays for both collider and SASE FEL proposals of TAC. Because of the high luminosity (~ $2.3\cdot10^{35}$ cm$^{-2}$s$^{-1}$) requirement of the TAC collider and high peak power (~2–3 GW) & high average brightness ($10^{25}$–$10^{30}$ photons/s/mrad$^2$/mm$^2$/0.1%BW) requirements of TAC SASE FEL, 1 GeV electron accelerator sector of the collider should be based on an Energy Recovery Linac [3], as shown in Fig. 6.

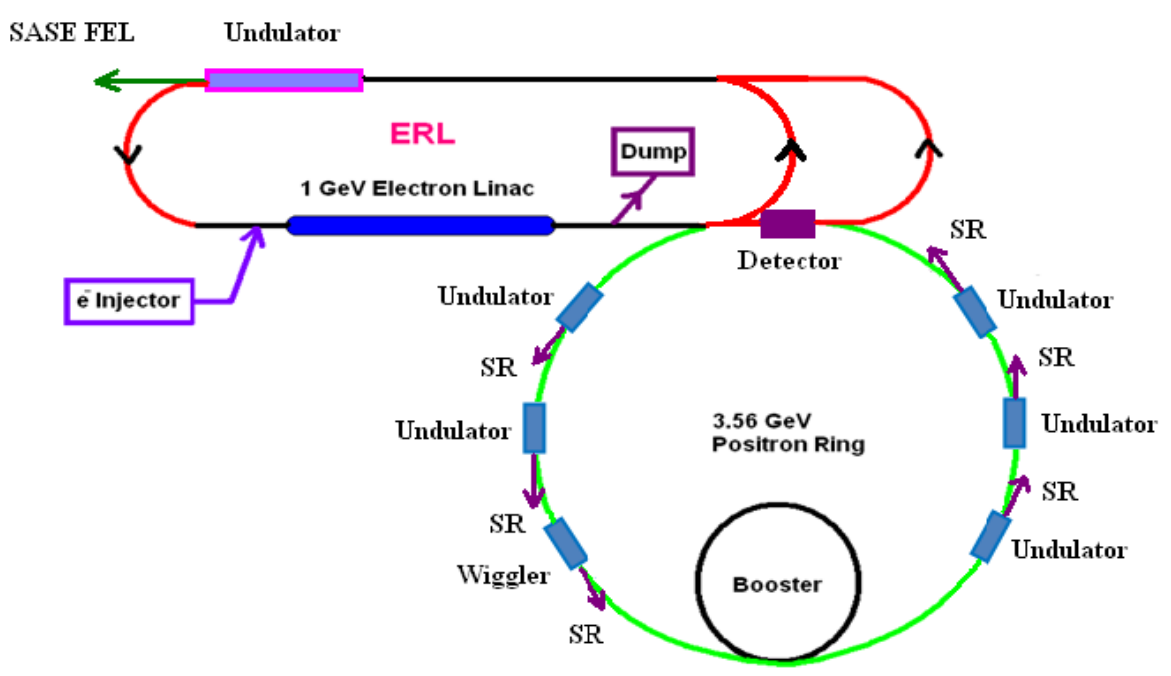

Figure 6: Schematic view of TAC ERL-Ring type $e^-e^+$ collider & SASE FEL facility

The modification on TAC collider from linac-ring type to ERL-ring type provides that, the total power consumption considerably decreases with a recovery about 95-98% for 1 GeV electron beam. This case directly lowers operating cost of the facility. Proposed TAC ERL & ring parameters [3] are given in Table 5. Asynchronous operation of the TAC collider & SASE FEL facility should be achieved by lowering the charge per bunch and arising the repetition rate in FEL operation, providing the same average current value for both collider and light source operations.

Table 5: TAC ERL-Ring $e^-e^+$ Parameters

| Parameter | cw $e^-$ ERL | $e^+$ Ring |
|---|---|---|
| Beam energy (GeV) | 1 | 3.56 |
| Num. of par. per bunch (x $10^{10}$) | 2 | 6 |
| Bunch charge (nC) | 3.2 | 9.6 |
| Bunch repetition rate (ns) | 6.67 | 6.67 |
| Bunch length (ps) | 20 | 20 |
| Number of bunches | - (cw) | 125 |
| RF frequency, $f_{RF}$ (MHz) | ~ $f_{coll}$ = 150 | ~ $f_{coll}$ = 150 |
| Rev. frequency, $f_{rev}$ (MHz) | - | 1.2 |
| Coll. frequency, $f_{coll}$ (MHz) | 150 | 150 |
| $\beta_x$ (mm) | 20 @ IP | 20 @ IP |
| $\beta_y$ (mm) | 0.5 @ IP | 0.5 @ IP |
| Normalized emittance, $\varepsilon_x$ (μm) | 3.92 | 14 |
| Normalized emittance, $\varepsilon_y$ (μm) | 0.06 | 0.2 |
| Transverse emittance, $\varepsilon_x$ (nm) | 2 | 2 |
| Transverse emittance, $\varepsilon_y$ (nm) | 0.03 | 0.03 |
| $\sigma_x$ (μm) | 6.32 | 6.32 |
| $\sigma_y$ (μm) | 0.12 | 0.12 |
| $\sigma_z$ (mm) | 6 | 6 |
| cw average current (A) | 0.48 | 1.44 |
| Peak current (kA) | 0.16 | 0.48 |

Also, the eRHIC proposal [4] is a promising reference for an asynchronously operating collider & light source from the World, which is based on a 5-10 GeV high current ERL.

## CONCLUSION

The TAC SASE FEL will act as a complementary part to cover the electromagnetic spectrum besides IR-FEL, Bremsstrahlung and Synchrotron Radiation proposals of TAC. Thereby, Turkish scientists and researchers will become acquainted with accelerator based light sources and its range of applications.

## ACKNOWLEDGEMENT

The authors would like to thank to all the contributors for this study on behalf of the TAC Project [5]. And also, special thanks to the Turkish State Planning Organization (SPO) for supporting the TAC project and to the staff for their complete assistance.

## REFERENCES

[1] Ciftci, A. K. *et al*. Linac-Ring Type Ø Factory for Basic and Applied Research, Turkish J. of Physics, 24 (2000), 747-758.
[2] General Design of SASE and Oscillator Mode Free Electron Lasers in Frame of the Turkish Accelerator Complex Project, S. Yigit, PhD. Thesis, Ankara University, 2007.
[3] Recepoglu, E., Sultansoy, S. A High Luminosity ERL on ring e-e+ collider for a super charm factory, e-Print: 0809.3233 [physics.acc-ph].
[4] Litvinenko, V.N., Ben-Zvi, I. Potential Use of eRHIC's ERL for FELs and Light Sources, Proceedings of FEL04, 594-597.
[5] http://thm.ankara.edu.tr